\documentstyle[11pt]{article}

\begin{document}

\topmargin 0pt \oddsidemargin 5mm 

\renewcommand{\thefootnote}{\fnsymbol{footnote}}

\newpage \setcounter{page}{0} 
\begin{titlepage}     
\begin{flushright}
UFSCAR-99-21
\end{flushright}
\vspace{0.5cm}
\begin{center}
\large{ Algebraic Bethe ansatz for the one-dimensional Hubbard model with
open boundaries} \\
\vspace{1cm}
\vspace{1cm}
 {\large X.-W. Guan } \\
\vspace{1cm}
\centerline{\em  Departamento de F\'isica, Universidade Federal de S\~ao Carlos}
\centerline{\em Caixa Postal 676, 13565-905, S\~ao Carlos, Brasil}
\centerline{ \em and }
\centerline{\em  Department of Physics, Jilin University}
\centerline{\em Changchun 130023, P. R. China}
\vspace{1.2cm}   
\end{center} 

\begin{abstract}
The one-dimensional Hubbard model with open boundary conditions is exactly
solved by means of algebraic Bethe ansatz. The eigenvalue of the transfer 
matrix, the energy spectrum as well as the Bethe ansatz equations are obtained.
\end{abstract}
\vspace{.2cm}
\centerline{PACS numbers: 71.27.+a, 75.10.Jm}
\vspace{.2cm}
\centerline{July 1999}
\end{titlepage}

\renewcommand{\thefootnote}{\arabic{footnote}}

\section{Introduction}

The one-dimensional (1D) Hubbard model has been one of the most fundamental
and favorite integrable models in the non-perturbative quantum field
theory. It exhibits on-site Coulomb interaction and correlated hopping which
will possiblely reveal a promising role in understanding the mystery of the
high-Tc superconductivity. Since Lieb and Wu \cite{LW}, in 1968, solved the
1D Hubbard model with periodic boundary conditions (BC) by coordinate Bethe
ansatz, there has been a great deal of papers devoted to the study of the
model. A remarkable step was done by Shastry \cite{Sha} who showed that the
Hamiltonian of the 1D Hubbard periodic chain commutes with a one-parameter
family of transfer matrix of an equivalent coupled symmetric $XY$ spin chain
and who also gave a direct proof of the integrability of the model by
presenting the quantum $R$-matrix. Later on, Wadati and coworkers \cite
{Wad1,Wad2} further studied its integrability in terms of quantum inverse
scattering method (QISM) \cite{QISM1,QISM2}. Very recently, Martins and
Ramos \cite{Mar} proposed a desirable way to solve the eigenvalue problem of
the transfer matrix of the 1D Hubbard model with periodic BC by means of
algebraic Beth ansatz. Their approach provides us a unified way to solve a
wide class of Hubbard-like models \cite{Hubl,zhou1} by algebraic Bethe
ansatz.

On the other hand, in recent years, there has been much interest in the
study of the quantum integrable systems with open BC, i.e. the systems on
finite interval with independent boundary conditions on each end. Due to the
presence of the boundary fields which lead to a pure back-scattering on each
end of the quantum chain and the exhibition of the quantum group symmetry by
special choice of the boundary parameters make the system possess rich
physical properties \cite{Frah,Kond} in thermodynamical point of view. A
systematic approach to handle the open BC for 1D integrable quantum chains
was proposed by Sklyanin \cite{SK}. A further extension of Sklyanin's
formalism to deal with more general class of models associated with Lie
(super) algebras was proposed by Mezincescu and Nepomechie \cite{MN}. We
also remark that the coordinate Bethe ansatz for 1D Hubbard model with
integrable boundary conditions was studied in \cite{Coor}. By now, though
there are several authors \cite{Hub1,Hub2,Hub3,Coor} have studied the open
BC for the 1D Hubbard model, the algebraic Bethe ansatz solution have not
yet been achieved. Actually, the diagonalization of the transfer matrix
which provide us with the spectrum of all conserved charges should be more
essential in studying finite temperature properties of the integrable models 
\cite{FTP,FTP2} than diagonalization of the underlying Hamiltonian. But as
we know the reflection equations for the 1D Hubbard model are much more
involved and the quantum $R$-matrix does not have the additive property that
make it difficult to built up the necessary commutation rules among the
diagonal fields and creation fields. In this paper, we intend to generalize
Sklyanin's formalism to solve the 1D Hubbard model with open BC. The
eigenvalue of the transfer matrix and Bethe ansatz equations for the model
will be given. It will be found that the model exhibits a hidden $XXX$ spin
open chain which play a crucial role to solve the model.

This paper is organized as follows. In section 2, we shall recall the main
results about open BC for the 1D Hubbard model in order to introduce the
notations which shall be used in this paper. In section 3, we perform the
algebraic Bethe ansatz approach for the model. In section 4, we formulate
the nested algebraic Bethe ansatz for the hidden $XXX$ quantum spin open
chain and present our main results. Section 5 is devoted to the conclusion.

\section{The 1D Hubbard model with boundary fields}

Let us consider the 1D Hubbard model with boundary fields determined by the
Hamiltonian \cite{Hub1,Hub2,Hub3} 
\begin{equation}
H= -\sum _{j=1}^{N-1}\sum _{s}(a^{\dagger }_{j+1s}a_{js}+a^{\dagger
}_{js}a_{j+1s}) +U\sum _{j=1}^{N}(n_{j\uparrow }-\frac{1}{2})(n_{j\downarrow
}-\frac{1}{2})  \label{2a}
\end{equation}
\[
+p_+(2n_{1\uparrow }-1)+p_-(2n_{1\downarrow }-1)+q_+(2n_{N\uparrow
}-1)+q_-(2n_{N\downarrow }-1).
\]
Here $p_{\pm }$ and $q_{\pm }$ are the free boundary parameters
characterizing the boundary fields. The coupling $U$ describes the on-site
Coulomb interaction and $a^{\dagger }_{js}$ and $a_{js}$ are creation and
annihilation operators with spins ($s=\uparrow $ or $\downarrow $) at site $j
$ satisfying the anti-commutation relations 
\begin{equation}
\{a_{js},a_{j^{^{\prime}}s^{^{\prime}}}\} = \{a^{\dagger }_{js},\,
a^{\dagger }_{j^{^{\prime}}s^{^{\prime}}}\}=0,\, \{a_{js},a^{\dagger
}_{j^{^{\prime}}s^{^{\prime}}}\} = \delta _{jj^{^{\prime}}}\delta
_{ss^{^{\prime}}},
\end{equation}
and $n_{js}=a^{\dagger }_{js}a_{js}$ is the density operator. The Lax
operator is given \cite{Wad1,Wad2} by 
\begin{equation}
{\cal {L}}_j(u)=\left(\matrix{-e^{h(u)}f_{j\uparrow
}f_{j\downarrow}&-f_{j\uparrow
}a_{j\downarrow}&ia_{j\uparrow}f_{j\downarrow}&ie^{h(u)}a_{j\uparrow}a_{j%
\downarrow}\cr -if_{j\uparrow }a^{\dagger
}_{j\downarrow}&e^{-h(u)}f_{j\uparrow
}g_{j\downarrow}&e^{-h(u)}a_{j\uparrow}a^{\dagger
}_{j\downarrow}&ia_{j\uparrow}g_{j\downarrow}\cr a^{\dagger }_{j\uparrow
}f_{j\downarrow}&e^{-h(u)}a^{\dagger }_{j\uparrow
}a_{j\downarrow}&e^{-h(u)}g_{j\uparrow}f_{j\downarrow}&g_{j\uparrow}a_{j%
\downarrow}\cr -ie^{h(u)}a^{\dagger }_{j\uparrow }a^{\dagger
}_{j\downarrow}&a^{\dagger }_{j\uparrow
}g_{j\downarrow}&ig_{j\uparrow}a^{\dagger
}_{j\downarrow}&-e^{h(u)}g_{j\uparrow}a_{j\downarrow}\cr }\right)  \label{2b}
\end{equation}
where 
\[
f_{js}=\sin u-(\sin u-i\cos u)n_{js},\,\, g_{js}=cos u-(\cos u+i\sin
u)n_{js}.
\]
With the grading $P(1)=P(4)=0, \, P(2)=P(3)=1$ and the constraint condition 
\begin{equation}
\frac{\sinh 2h(u)}{\sin 2u}=\frac{U}{4},
\end{equation}
the Lax operator (\ref{2b}) satisfies the graded Yang-Baxter algebra 
\begin{equation}
{\cal R}_{12}(u,v)\stackrel{1}{{\cal T}}(u)\stackrel{2}{{\cal T}}(v)=%
\stackrel{2}{{\cal T}}(v)\stackrel{1}{{\cal T}}(u){\cal R}_{12}(u,v),
\label{2c}
\end{equation}
where the monodromy matrix ${\cal {T}}(u)$ is defined by 
\begin{equation}
{\cal {T}}(u)={\cal {L}}_N(u)\cdots {\cal {L}}_1(u),  \label{2d}
\end{equation}
and 
\begin{equation}
\stackrel{1}{{\cal T}}(u)={\cal T} (u)\otimes _{s} I;\, \stackrel{2}{{\cal T}%
}(u)=I\otimes _{s}{\cal T}(u).
\end{equation}
here ${\otimes }_{s}$ is the super direct product: 
\[
[A{\otimes }_{s}B]_{\alpha \beta , \gamma \delta}=(-1)^{[P(\alpha )+P(\gamma
)]P(\beta )}A_{\alpha \gamma }B_{\beta \delta}.
\]
For our convenience in practical calculation, we display the associated
quantum ${\cal R}_{12}(u,v)$-matrix in Appendix. One may show that ${\cal R}%
_{12}(u,v)$ enjoys the following graded reflection equations \cite{Hub3} : 
\begin{equation}
{\cal R}_{12}(u,v)\stackrel{1}{K_-}(u){\cal R}_{21}(v,-u)\stackrel{2}{K_-}%
(v)= \stackrel{2}{K_-}(v){\cal R}_{12}(u,-v)\stackrel{1}{K_-}(u){\cal R}%
_{21}(-v,-u),  \label{2e}
\end{equation}

\begin{eqnarray}
{\cal R}_{21}^{{\rm St}_1\stackrel{-}{{\rm St}}_2}(v,u)\stackrel{1}{K_+^{%
{\rm St}_1}}(u){\cal R}_{12}^{{\rm St}_1\stackrel{-}{{\rm St}}_2}(-u,v) 
\stackrel{2}{K_+^{\stackrel{-}{{\rm St}}_2}}(v)=  \nonumber \\
\stackrel{2}{K_+^{\stackrel{-}{{\rm St}}_2}}(v){\cal R}_{21}^{{\rm St}_1 
\stackrel{-}{{\rm St}}_2}(-v,u)\stackrel{1}{K_+^{{\rm St}_1}}(u) {\cal R}%
_{12}^{{\rm St}_1\stackrel{-}{{\rm St}}_2}(-u,-v)  \label{2e2}
\end{eqnarray}

which ensure the integrability of the model (\ref{2a}) provided that 
\begin{equation}
K_{\pm }(u)=\left(\matrix{K1_{\pm }(u)&0&0&0\cr 0&K2_{\pm }(u)&0&0\cr
0&0&K3_{\pm }(u)&0\cr 0&0&0&K4_{\pm }(u)\cr}\right),  \label{2f}
\end{equation}
where $p_+=p_-=\xi _-/2$

\begin{eqnarray}
K1_-(u) & = & \lambda _-(e^{-h(u)}\cos u-e^{h(u)}\xi _-\sin u)(e^{h(u)}\cos
u-e^{-h(u)}\xi _-\sin u),  \nonumber \\
K2_-(u) & = & \lambda _-(e^{-h(u)}\cos u+e^{h(u)}\xi _-\sin u)(e^{-h(u)}\cos
u-e^{h(u)}\xi _-\sin u),  \nonumber \\
K3_-(u) & = & \lambda _-(e^{-h(u)}\cos u+e^{h(u)}\xi _-\sin u)(e^{-h(u)}\cos
u-e^{h(u)}\xi _-\sin u),  \label{km} \\
K4_-(u) & = & \lambda _-(e^{h(u)}\cos u+e^{-h(u)}\xi _-\sin u)(e^{-h(u)}\cos
u+e^{h(u)}\xi _-\sin u),  \nonumber
\end{eqnarray}
and $q_+=q_-=\xi _+/2$

\begin{eqnarray}
K1_{+}(u) &=&\lambda _{+}(e^{-h(u)}\xi _{+}\cos u+e^{h(u)}\sin
u)(e^{h(u)}\xi _{+}\cos u+e^{-h(u)}\sin u),  \nonumber \\
K2_{+}(u) &=&\lambda _{+}(e^{h(u)}\xi _{+}\cos u+e^{-h(u)}\sin
u)(e^{h(u)}\xi _{+}\cos u-e^{-h(u)}\sin u),  \nonumber \\
K3_{+}(u) &=&\lambda _{+}(e^{h(u)}\xi _{+}\cos u+e^{-h(u)}\sin
u)(e^{h(u)}\xi _{+}\cos u-e^{-h(u)}\sin u),  \label{kp} \\
K4_{+}(u) &=&\lambda _{+}(e^{h(u)}\xi _{+}\cos u-e^{-h(u)}\sin
u)(e^{-h(u)}\xi _{+}\cos u-e^{h(u)}\sin u).  \nonumber
\end{eqnarray}
Here $\lambda _{\pm }$ and $\xi _{\pm }$ are arbitary constants describing
boundary effects. $\stackrel{-}{{\rm St}}_{a}$ stands for the inverse of the
supertransposition in the space $a$. The supertransposition is defined by 
\[
(A_{ij})^{{\rm St}}=(-1)^{[P(i)+1]P(j)}A_{ji}.
\]
We would like to remark that Zhou \cite{Hub1} first time gave a class of
boundary $K_{\pm }$-matrices equivalent to (\ref{km}) and (\ref{kp}) in
terms of QISM. Consequently, using Lax pair formulation, the author \cite
{Hub2} presented two class of boundary $K_{\pm }$-matrices leading to four
possible boundary terms in the 1D Hubbard open chain Hamiltonian. While,
Shiroishi and Wadati \cite{Hub3} studied the open BC for the model in terms
of the graded version of QISM and also presented two class of the solutions
to the graded RE. The second solution to the graded RE (\ref{2e}) and (\ref
{2e2}) permits the boundary fields with $p_{+}=-p_{-}$ and $q_{+}=-q_{-}$
corresponding to magnetic boundary fields (see ref. \cite{Hub2,Hub3} ). In
this paper, we restrict to study the chemical boundary fields (\ref{km}) and
(\ref{kp}) basing on the consideration that this kind of boundary conditions
will bring us a simple boundary $K$-matrix for the hidden $XXX$ open chain.
To other kinds of boundary conditions, of course, we may treat them in a
similar way. It is found that the Hamiltonian (\ref{2a}) is related to the
double-row monodromy matrix 
\begin{equation}
\tau (u)={\rm Str}_{0}K_{+}(u){\cal T}(u)K_{-}(u){\cal T}^{-1}(-u)
\label{2h}
\end{equation}
in the following way 
\begin{equation}
\tau (u)=c_{1}u+c_{2}u^{2}+c_{3}(H+{\rm const.})~u^{3}+\cdots   \label{2g}
\end{equation}
where $c_{i},\,i=1,\cdots 4,$ are some scalar functions of boundary
parameters. ${\rm Str}_{0}$ denotes the supertrace carried out in auxiliary
space $v_{0}$.

\section{Algebraic Bethe ansatz approach}

According to the algebraic Bethe ansatz, let us first choose the standard
ferromagnetic pseudovacuum state $|0\rangle_i$ 
\begin{equation}
|0\rangle_i=\left(
\begin{array}{c}
1 \\ 
0
\end{array}
\right)_i\otimes _s\left(
\begin{array}{c}
1 \\ 
0
\end{array}
\right)_i
\end{equation}
as a highest vector, which corresponds to the doubly occupied state.
Following the notation introduced in \cite{Ram}, we define the monodromy
matix ${\cal T}(u)$ as 
\begin{eqnarray}
{\cal T}(u) & = & \left(\matrix{B(u)&B_1(u)&B_2(u)&F(u)\cr
C_1(u)&A_{11}(u)&A_{12}(u)&E_1(u)\cr C_2(u)&A_{21}(u)&A_{22}(u)&E_2(u)\cr
C_3(u)&C_4(u)&C_5(u)&D(u)\cr }\right) \\
{\cal T}^{-1}(-u) & = & \left(\matrix{\stackrel{-}{B}(u)&%
\stackrel{-}{B}_1(u)&\stackrel{-}{B}_2(u)&\stackrel{-}{F}(u)\cr
\stackrel{-}{C}_1(u)&\stackrel{-}{A}_{11}(u)&\stackrel{-}{A}_{12}(u)&%
\stackrel{-}{E}_1(u)\cr
\stackrel{-}{C}_2(u)&\stackrel{-}{A}_{21}(u)&\stackrel{-}{A}_{22}(u)&%
\stackrel{-}{E}_2(u)\cr \stackrel{-}{
C}_3(u)&\stackrel{-}{C}_4(u)&\stackrel{-}{C}_5(u)&\stackrel{-}{D}(u)\cr }%
\right) \\
\end{eqnarray}
and 
\begin{eqnarray}
{\cal T}_-(u) & = & {\cal T}(u)K_-(u){\cal T}^{-1}(-u)  \nonumber \\
& = & \left(\matrix{\tilde{B}(u)&\tilde{B}_1(u)&\tilde{B}_2(u)&\tilde{F}(u)%
\cr \tilde{C}_1(u)&\tilde{A}_{11}(u)&\tilde{A}_{12}(u)&\tilde{E}_1(u)\cr
\tilde{C}_2(u)&\tilde{A}_{21}(u)&\tilde{A}_{22}(u)&\tilde{E}_2(u)\cr
\tilde{C}_3(u)&\tilde{C}_4(u)&\tilde{C}_5(u)&\tilde{D}(u)\cr }\right),
\label{3a}
\end{eqnarray}
It is not difficult to show that ${\cal T}_-(u)$ also satisfies the RE (\ref
{2e}). Acting ${\cal T}(u)$ and ${\cal T}^{-1}(-u)$ on the pseudovacuum
state 
\begin{equation}
|0\rangle=\otimes _{i=1}^N|0\rangle_i,
\end{equation}
we have the following properties (upon a common factor): 
\begin{eqnarray}
B(u) |0\rangle & = & \stackrel{-}{B}(u)|0\rangle=\left\{\frac{\cos u}{\sin u}%
e^{2h(u)}\right\}^N|0\rangle,  \nonumber \\
D(u) |0\rangle & = & \stackrel{-}{D}(u)|0\rangle=\left\{\frac{\sin u}{\cos u}%
e^{2h(u)}\right\}^N|0\rangle,  \nonumber \\
A_{aa}(u) |0\rangle & = & \stackrel{-}{A}_{aa}(u)|0\rangle=|0\rangle, \, a=1,2, 
\nonumber \\
A_{ab}(u) |0\rangle & = & \stackrel{-}{A}_{ab}(u)|0\rangle=0, \, a\neq b=1,2, 
\nonumber \\
B_a(u) |0\rangle & \neq & 0,\, \stackrel{-}{B}_a(u)|0\rangle \neq 0, \, a=1,2,
\label{3bb} \\
E_a(u) |0\rangle & \neq & 0,\, \stackrel{-}{E}_a(u)|0\rangle \neq 0, \, a=1,2, 
\nonumber \\
F(u) |0\rangle & \neq & 0,\, \stackrel{-}{F}(u)|0\rangle \neq 0,  \nonumber \\
C_i(u) |0\rangle & = & \stackrel{-}{C}_i(u)|0\rangle = 0, \, i=1,\cdots ,5. 
\nonumber
\end{eqnarray}
Using the properties (\ref{3bb}), and the Yang-Baxter algebra 
\begin{equation}
\stackrel{2}{{\cal T}}^{-1}(-u){\cal R}_{12}(u,-u)\stackrel{1}{{\cal T}}(u)= 
\stackrel{1}{{\cal T}}(u){\cal R}_{12}(u,-u)\stackrel{2}{{\cal T}}^{-1}(-u),
\end{equation}
and after some algebra, one can obtain 
\begin{eqnarray}
\tilde{B}(u)|0\rangle & = & W_1^-(u)B(u)\stackrel{-}{B}(u)|0\rangle, \\
\tilde{A}_{aa}(u)|0\rangle & = & \{\frac{\rho _2(u,-u)}{\rho _1(u,-u)}B(u)%
\stackrel{-}{B}(u) +W_2^-(u)A_{aa}(u)\stackrel{-}{A}_{aa}(u)\}|0\rangle, \\
\tilde{D}(u)|0\rangle & = & \left\{\frac{1}{\rho _4(u,-u)}(K2_-(u)- \frac{\rho
_2(u,-u)}{\rho _1(u,-u)})\sum_{a=1}^2A_{aa}(u)\stackrel{-}{A}_{aa}(u) \right.
\nonumber \\
& & \left.+\frac{\rho _3(u,-u)}{\rho _1(u,-u)}B(u)\stackrel{-}{B}%
(u)+W^-_4(u)D(u)\stackrel{-}{D}(u)\right\}|0\rangle, \\
\tilde{B}_a(u) |0\rangle & \neq & 0,\, \tilde{E}_a(u) |0\rangle \neq 0, \, a=1,2,
\\
\tilde{A}_{ab}(u) |0\rangle & = & 0, \, a\neq b=1,2,\, \tilde{F}(u) \neq 0, \\
\tilde{C}_i(u) |0\rangle & = & 0 ,\, i=1,\cdots ,5,
\end{eqnarray}
where

\begin{eqnarray}
W^-_1(u) & = & 1, \\
W_2^-(u) & = & -\frac{(e^{-2h(u)}+e^{2h(u)})\sin u\cos u(\xi _-e^{h(u)}\cos
u-e^{-h(u)}\sin u)} {(e^{2h(u)}\cos^2 u-e^{-2h(u)}\sin ^2u)(\xi
_-e^{-h(u)}\sin u-e^{h(u)}\cos u)}, \\
W_4^-(u) & = & \frac{(e^{-2h(u)}+e^{2h(u)})\sin u\cos u\sin 2u }{\cos2u
(e^{-2h(u)}\cos^2 u-e^{2h(u)}\sin ^2u)}\times  \nonumber \\
& & \frac{(e^{-h(u)}\xi _-\cos u-e^{h(u)}\sin u)(e^{h(u)}\xi _-\cos
u+e^{-h(u)}\sin u)} {(e^{-h(u)}\cos u-e^{h(u)}\xi _-\sin u)(e^{h(u)}\cos
u-e^{-h(u)}\xi _-\sin u)}.
\end{eqnarray}
We also notice that the operators $\tilde{B}_a(u)$, $\tilde{E}_a(u)$ and $\tilde{F}(u)$ are still creation fields, otherwise, $\tilde{C}_i(u)$ are the annihilation fields. Via the transformations

\begin{eqnarray}
\tilde{A}_{aa}^{^{\prime}}(u) & = & \tilde{A}_{aa}(u)-\frac{\rho _2(u,-u)}{%
\rho _1(u,-u)}\tilde{B}(u), \\
\tilde{D}^{^{\prime}}(u) & = & \tilde{D}(u)-\frac{\rho _3(u,-u)}{\rho
_1(u,-u)}\tilde{B}(u)- \frac{1}{\rho _4(u,-u)}\sum_{a=1}^2\tilde{A}%
_{aa}^{^{\prime}}(u),
\end{eqnarray}
we may express the transfer matrix (\ref{2h}) in the following way 
\begin{eqnarray}
\tau(u) & = & {\rm Str}_0K_+(u){\cal T}_-(u)  \nonumber \\
& = & W^+_1(u)\tilde{B}(u)+W^+_2(u)\sum_{a=1}^2\tilde{A}_{aa}^{^{%
\prime}}(u)+W^+_4(u)\tilde{D}^{^{\prime}}(u),  \label{3b}
\end{eqnarray}
where 
\begin{eqnarray}
W^+_1(u) & = & \frac{(e^{-2h(u)}+e^{2h(u)})\sin u\cos u\sin 2u}{\cos
2u(e^{2h(u)}\cos^2 u-e^{-2h(u)}\sin ^2u)}f(u)\times  \nonumber \\
& & \frac{(e^{-h(u)}\xi _+\sin u-e^{h(u)}\cos u)(e^{h(u)}\xi _+\sin
u-e^{-h(u)}\cos u)} {(e^{-h(u)}\xi _+\cos u+e^{h(u)}\sin u)(e^{h(u)}\xi
_+\cos u+e^{-h(u)}\sin u)}, \\
W^+_2(u) & = & \frac{(e^{-2h(u)}+e^{2h(u)})\sin u\cos u}{(e^{-2h(u)}\cos^2
u-e^{2h(u)}\sin ^2u)}f(u)\times  \nonumber \\
& & \frac{(e^{h(u)}\xi _+\cos u-e^{-h(u)}\sin u)(e^{h(u)}\xi _+\sin
u-e^{-h(u)}\cos u)} {(e^{-h(u)}\xi _+\cos u+e^{h(u)}\sin u)(e^{h(u)}\xi
_+\cos u+e^{-h(u)}\sin u)} \\
W_4^+ & = & \frac{(e^{-h(u)}\xi _+\cos u-e^{h(u)}\sin u)(e^{h(u)}\xi _+\cos
u-e^{-h(u)}\sin u)} {(e^{-h(u)}\xi _+\cos u+e^{h(u)}\sin u)(e^{h(u)}\xi
_+\cos u+e^{-h(u)}\sin u)}f(u)
\end{eqnarray}
with 
\begin{equation}
f(u)=e^{-2Nh(u)}\cos ^{2N}u~\sin ^{2N}u~K1_-(u)K1_+(u).  \label{fcom}
\end{equation}

Now we proceed the key step to built up the necessary commutation relations
between the diagonal fields and the creation fields respectively. From the
RE (\ref{2e}) and definition (\ref{3a}), after many steps of substitution,
we can get the following important commutation relations: 
\begin{eqnarray}
\tilde{B}(u)\tilde{B}_a(v) & = & \frac{\rho _1(v,u)\rho _{10}(u,-v)}{\rho
_1(v,-u)\rho _{10}(-u,-v)}\tilde{B}_a(v)\tilde{B}(u)+u.t.,  \label{3c} \\
\tilde{D}^{^{\prime}}(u)\tilde{B}_a(v) & = & -\frac{\rho _7(u,-v)\rho
_{9}(-v,-u)}{\rho _9(u,-v)\rho _{8}(u,v)}\tilde{B}_a(v)\tilde{D}%
^{^{\prime}}(u)+u.t.,  \label{3d} \\
\tilde{A}^{^{\prime}}_{ab}(u)\tilde{B}_a(v) & = & -\frac{\rho _4(-v,-u)\rho
_{10}(u,-v)}{\rho _1(u,-v)\rho _{9}(u,v)}r^{ea}_{gh}(u,-v)\stackrel{-}{r}%
^{ih}_{cb}(-v,-u)\tilde{B}_e(v)\tilde{A}^{^{\prime}}_{gi}(u)+u.t.,
\label{3e} \\
\tilde{B}_a(u)\otimes \tilde{B}_b(v) & = & \frac{\rho _{10}(u,-v)\rho
_{4}(-v,-u)}{\rho _1(u,v)\rho _{10}(v,-u)}\left\{\tilde{B}_c(v)\otimes 
\tilde{B}_d(u)\right.  \nonumber \\
& & \left.-\frac{\rho _6(u,-v)}{\rho _{10}(u,-v)}F(v)\stackrel{\rightarrow}{%
\xi }(I\otimes \hat{A}(u))\right\}.\stackrel{-}{r}(-v,-u)  \nonumber \\
& & +\frac{\rho _6(v,-u)}{\rho _{10}(v,-u)}F(u)\stackrel{\rightarrow}{\xi }%
(I\otimes \hat{A}(v))  \nonumber \\
& & +\frac{\rho _{8}(v,-u)\rho _{6}(-v,-u)}{\rho _{10}(v,-u)\rho _{8}(-v,-u)%
}[\tilde{F}(v)\tilde{B}(u)-\tilde{F}(u)\tilde{B}(v)].\stackrel{\rightarrow}{%
\xi },  \label{3f}
\end{eqnarray}
where 
\[
\stackrel{\rightarrow }{\xi }=(0,1,-1,0), 
\]
\begin{equation}
r(u,-v) = \left(\matrix{1&0&0&0\cr 0&a(u,-v)&b(u,-v)&0\cr
0&b(u,-v)&a(u,-v)&0 \cr 0&0&0&1\cr }\right),  \label{r1}
\end{equation}
\begin{equation}
\stackrel{-}{r}(-v,-u) = \left(\matrix{1&0&0&0\cr
0&\stackrel{-}{a}(-v,-u)&\stackrel{-}{b}(-v,-u)&0\cr
0&\stackrel{-}{b}(-v,-u)&\stackrel{-}{a}(-v,-u)&0 \cr 0&0&0&1\cr }\right),
\label{r2}
\end{equation}
with the weights 
\begin{eqnarray}
a(u,-v) & = & \frac{\rho _3(u,-v)\rho _4(u,-v)-1}{\rho _9(u,-v)\rho
_{10}(u,-v)},\, b(u,-v)=1-a(u,-v), \\
\stackrel{-}{a}(-v,-u) & = & \frac{\rho _5(-v,-u)\rho _8(-v,-u)+\rho
_6^2(-v,-u)}{\rho _4(-v,-u)\rho _8(-v,-u)},\, \stackrel{-}{b}(-v,-u)=1-%
\stackrel{-}{a}(-v,-u).
\end{eqnarray}
In the commutation relations (\ref{3c})-(\ref{3e}), we had to omit all
unwanted terms because they take a big space to display. It turns out that
the auxiliary matrices $r(u,-v)$ and $\stackrel{-}{r}(-v,-u)$ are nothing
but the rational $R$-matrices of isotropic six-vertex model. The structure
of the auxiliary matrics is very important to solve the Hubbard-like \cite
{zhou1,SLMG} models with open BC that exhibit a similar structure of the
auxiliary matrices like Eq. (\ref{r1}) and (\ref{r2}). If performing the
parameterization introduced in \cite{Mar}, 
\begin{equation}
\tilde x=\frac{\sin x}{\cos x}e^{-2h(x)}-\frac{\cos x}{\sin x}%
e^{2h(x)},x=u,v,
\end{equation}
one may also find 
\begin{eqnarray}
a(u,-v) & = & \frac{U}{\tilde{u}+\tilde{v}+U},\, b(u,-v)=\frac{\tilde{u}+%
\tilde{v}}{\tilde{u}+\tilde{v}+U}, \\
\stackrel{-}{a}(-v,-u) & = & \frac{U}{\tilde{u}-\tilde{v}+U},\,\stackrel{-}{b%
}(-v,-u)=\frac{\tilde{u}-\tilde{v}}{\tilde{u}-\tilde{v}+U}.
\end{eqnarray}
In view of the commutation relation (\ref{3f}), the creation operators $%
\tilde{B}_a,\tilde{E}_a$ do not interwine. So it is reasonable that the
eigenvectors of the transfer matrices are generated only by the creation
operators $B_a(u)$ and $F(u)$ or $E_a(u)$ and $F(u)$. Unfortunately, it
seems to be very difficult to construct the explicit form of the
multi-particle vector even in the case of the Hubbard periodic chain \cite
{Mar}. But it does have a similar recursive relation as that for the Hubbard
periodic chain. Here we prefer the $n$-particle vector in a formal form,
namely 
\begin{equation}
\mid \Phi _n(v_1,\cdots,v_n)\rangle=\Phi _{n}(v_1,\cdots,v_n)F^{\mbox{\scriptsize
$a_1,\cdots ,a_n$}}|0\rangle.  \label{3g}
\end{equation}
Where the $n$-particle vector $\Phi _{n}(v_1,\cdots,v_n)$ may be given by a
recursive relation 
\begin{eqnarray}
\Phi _{n}(v_1,\cdots,v_n) & = & \tilde{B}_a(v_1)\otimes\Phi
_{n-1}(v_2,\cdots,v_n)  \nonumber \\
& & +\sum _{j=2}^n[\stackrel{\rightarrow }{\xi }\otimes \tilde{F}(v_1)] \Phi
_{n-2}(v_2,\cdots,v_{j-1},v_{j+1},\cdots,v_n)\tilde{B}%
(v_j)g^{(n)}_{j-1}(v_1,\cdots,v_n) \\
& & -\sum _{j=2}^n[\stackrel{\rightarrow }{\xi }\otimes \tilde{F}(v_1)] \Phi
_{n-2}(v_2,\cdots,v_{j-1},v_{j+1},\cdots,v_n)(I\otimes \hat{A}%
(u))h^{(n)}_{j-1}(v_1,\cdots,v_n).  \nonumber
\end{eqnarray}
From the commutation relation (\ref{3f}), we can conclude that $\Phi
_{n}(v_1,\cdots,v_n)$ also satisfies the symmetry relation 
\begin{eqnarray}
\Phi _{n}(v_1,\cdots,v_j,v_{j+1},\cdots,v_n) & = & \frac{\rho
_{10}(v_j,-v_{j+1})\rho _{4}(-v_{j+1},-v_j)}{\rho _1(v_j,v_{j+1})\rho
_{10}(v_{j+1},-v_j)}\times  \nonumber \\
& & \Phi _{n}(v_1,\cdots,v_{j+1},v_{j},\cdots,v_n).\stackrel{-}{r}%
(-v_{j+1},-v_{j})
\end{eqnarray}
basing on the following relation: 
\begin{equation}
\frac{\rho _{4}(-v_{j+1},-v_j)}{\rho _1(v_j,v_{j+1})} \stackrel{\rightarrow 
}{\xi }.\stackrel{-}{r}(-v_{j+1},-v_{j})= \frac{\rho
_{8}(v_{j+1},-v_{j})\rho _{6}(-v_{j+1},-v_j)\rho _{8}(-v_{j},-v_{j+1})} {%
\rho _8(-v_{j+1},v_{j})\rho _{8}(v_{j},-v_{j+1})\rho _{6}(-v_{j},-v_{j+1})}.%
\stackrel{\rightarrow }{\xi }.
\end{equation}
This symmetry giving a restriction to the functions $h^{(n)}_{j-1}(v_1,%
\cdots,v_n)$ and $g^{(n)}_{j-1}(v_1,\cdots,v_n)$ is very useful to deduce
the coefficients and simplify the unwanted terms in the eigenvalue of the
transfer matrix. In fact, after performing three-particle scattering, the
explicit form of these coefficients can be fixed. But checking
three-particle scattering is indeed a extremely tough work, we had to leave
the coefficients to be determined. Explicitly, we display the two-particle
state 
\begin{eqnarray}
\Phi _{2}(v_1,v_2) & = & \tilde{B}_{\mbox{\scriptsize $a_1$}}(v_1)\otimes 
\tilde{B}_{\mbox{\scriptsize $a_2$}}(v_2) -\frac{\rho _6(v_2,-v_1)}{\rho
_{10}(v_2,-v_1)}F(v_1)\stackrel{\rightarrow}{\xi }(I\otimes \hat{A}(v_2)) 
\nonumber \\
& & +\frac{\rho _{8}(v_2,-v_1)\rho _{6}(-v_2,-v_1)}{\rho _{10}(v_2,-v_1)\rho
_{8}(-v_2,-v_1)}\tilde{F}(v_1)\tilde{B}(v_2).\stackrel{\rightarrow}{\xi }.
\end{eqnarray}
In above expressions, $F^{\mbox{\scriptsize $a_1,\cdots ,a_n$}}$ are the
coefficients of arbitrary linear combination of the vectors reflect the
"spin" degrees of freedom with $a_i=1,2$ and $\stackrel{\rightarrow }{\xi }$
plays the role of forbidding two spin up or two spin down at same site. $%
\tilde{F}(u)$ creats a local hole pair with opposite spins. Acting the
diagonal fields on the vector (\ref{3g}), we way phenomenologically get 
\begin{eqnarray}
\tilde{B}(u)\mid \Phi _n(v_1,\cdots,v_n)\rangle & = & \tilde{B}(u)\prod _{i=1}^n%
\frac{\rho _1(v_i,u)\rho _{10}(u,-v_i)}{\rho _1(v_i,-u)\rho _{10}(-u,-v_i)}%
\mid \Phi _n(v_1,\cdots,v_n)\rangle+u.t., \\
\tilde{D}(u)\mid \Phi _n(v_1,\cdots,v_n)\rangle & = & \tilde{D}^{^{\prime}}(u)%
\prod _{i=1}^n-\frac{\rho _7(u,-v_i)\rho _{9}(-v_i,-u)}{\rho _9(u,-v_i)\rho
_{8}(u,v_i)}\mid \Phi _n(v_1,\cdots,v_n)\rangle+u.t., \\
\tilde{A}_{aa}^{^{\prime}}(u)\mid \Phi _n(v_1,\cdots,v_n)\rangle & = & \tilde{A}%
_{aa}^{^{\prime}}(u)\prod _{i=1}^n-\frac{\rho _4(-v_i,-u)\rho _{10}(u,-v_i)}{%
\rho _1(u,-v_i)\rho _{9}(u,v_i)}r_{12}(\tilde{u}+\tilde{v}_1)_{%
\mbox{\scriptsize $h_1g_1$}}^{\mbox{\scriptsize $e_1a$}} \times  \nonumber \\
& & r_{12}(\tilde{u}-\tilde{v}_1)_{\mbox{\scriptsize $al_1$}}^{%
\mbox{\scriptsize $i_1h_1$}} r_{12}(\tilde{u}+\tilde{v}_2)_{%
\mbox{\scriptsize $h_2g_2$}}^{\mbox{\scriptsize $e_2g_1$}}r_{12}(\tilde{u}-%
\tilde{v}_2)_{\mbox{\scriptsize $i_1l_2$}}^{\mbox{\scriptsize $i_2h_2$}}
\cdots  \nonumber \\
& & r_{12}(\tilde{u}+\tilde{v}_n)_{\mbox{\scriptsize $h_ng_n$}}^{%
\mbox{\scriptsize $e_ng_{n-1}$}}r_{12}(\tilde{u}-\tilde{v}_n)_{%
\mbox{\scriptsize $i_{n-1}l_n$}}^{\mbox{\scriptsize $i_nh_n$}} \mid \Phi
_n(v_1,\cdots,v_n)\rangle+u.t.
\end{eqnarray}
It follows that 
\begin{eqnarray}
\tau(u)\mid \Phi _n(v_1,\cdots,v_n)\rangle & = & \left\{W_1^+(u)\tilde{B}(u)\prod
_{i=1}^n\frac{\rho _1(v_i,u)\rho _{10}(u,-v_i)}{\rho _1(v_i,-u)\rho
_{10}(-u,-v_i)}\right.  \nonumber \\
& & \left.+W^+_4(u)\tilde{D}^{^{\prime}}(u)\prod _{i=1}^n-\frac{\rho
_7(u,-v_i)\rho _{9}(-v_i,-u)}{\rho _9(u,-v_i)\rho _{8}(u,v_i)}\right. 
\nonumber \\
& & \left.+W^+_2(u)\tilde{A}_{aa}^{^{\prime}}(u)\prod _{i=1}^n-\frac{\rho
_4(-v_i,-u)\rho _{10}(u,-v_i)}{\rho _1(u,-v_i)\rho _{9}(u,v_i)} \Lambda
^{(1)}(\tilde{u},\{\tilde{v}_i\})\right\}\times  \nonumber \\
& & \mid \Phi _n(v_1,\cdots,v_n)\rangle
\end{eqnarray}
provided that 
\begin{equation}
\frac{W^+_1(u)\tilde{B}(u)}{W^+_2(u)\tilde{A}^{^{\prime}}_{11}(u)}\mid _{%
\mbox{\scriptsize $u=v_i$}}=\Lambda ^{(1)}(\tilde{u}=\tilde{v}_i,\{\tilde{v}%
_i\}).
\end{equation}
\[
i=1,\cdots n.
\]
Here $r_{12}(u)=P.r(u)$ and $\Lambda ^{(1)}(\tilde{u},\{\tilde{v}_i\})$ is
the eigenvalue of the nested transfer matrix (\ref{3i}), i.e. 
\begin{equation}
\tau^{(1)}(\tilde{u},\{\tilde{v}_i\})F^{\mbox{\scriptsize $e_1\cdots e_n$}%
}=\Lambda ^{(1)}(\tilde{u},\{\tilde{v}_i\})F^{\mbox{\scriptsize $e_1\cdots
e_n$}},  \label{3h}
\end{equation}
where 
\begin{equation}
\tau^{(1)}(\tilde{u},\{\tilde{v}_i\})={\rm Tr}_0T^{(1)}(\tilde{u}){T^{(1)}}%
^{-1}(-\tilde{u}).  \label{3i}
\end{equation}
The nested monodromy matrices $T^{(1)}(\tilde{u})$ and ${T^{(1)}}^{-1}(-%
\tilde{u})$ read 
\begin{eqnarray}
T^{(1)}(\tilde{u}) & = & r_{12}(\tilde{u}+\tilde{v}_1)_{\mbox{\scriptsize
$h_1g_1$}}^{\mbox{\scriptsize $e_1a$}}\cdots r_{12}(\tilde{u}+\tilde{v}_n)_{%
\mbox{\scriptsize $h_ng_n$}}^{\mbox{\scriptsize $e_ng_{n-1}$}}, \\
{T^{(1)}}^{-1}(-\tilde{u}) & = & r_{12}(\tilde{u}-\tilde{v}_1)_{%
\mbox{\scriptsize $i_{n-1}l_n$}}^{\mbox{\scriptsize $i_nh_n$}}\cdots r_{12}(%
\tilde{u}-\tilde{v}_1)_{\mbox{\scriptsize $al_1$}}^{\mbox{\scriptsize
$i_1h_1$}}.
\end{eqnarray}
So far, the eigenvalue problem of the 1D Hubbard model with boundaries
reduces to solve the nested auxiliary transfer matrix (\ref{3h}) which
corresponds to an isotropic six-vertex model with open boundary conditions.

\section{The nested Bethe ansatz}

In this section, we proceed the diagonalization of the auxiliary transfer
matrix (\ref{3i}). Following Sklyanin's formalism \cite{SK}, performing the
nested Bethe ansatz have not been a difficult problem yet. It is easy to
check that the $r_{12}(u)$-matrix satisfies the Yang-Baxter algebra 
\begin{equation}
r_{12}(\tilde{u}_{1}-\tilde{u}_{2})\stackrel{1}{T^{(1)}}(\tilde{u}_{1},\{%
\tilde{v}_{i}\})\stackrel{2}{T^{(1)}}(\tilde{u}_{2},\{\tilde{v}_{i}\})=%
\stackrel{2}{T^{(1)}}(\tilde{u}_{2},\{\tilde{v}_{i}\})\stackrel{1}{T^{(1)}}(%
\tilde{u}_{1},\{\tilde{v}_{i}\})r_{12}(\tilde{u}_{1}-\tilde{u}_{2}),
\label{4a}
\end{equation}
and the reflection equations 
\begin{equation}
r_{12}(\tilde{u}_{1}-\tilde{u}_{2})\stackrel{1}{K_{-}^{(1)}}(\tilde{u}%
_{1})r_{12}(\tilde{u}_{1}+\tilde{u}_{2})\stackrel{2}{K_{-}^{(1)}}(\tilde{u}%
_{2})=\stackrel{2}{K_{-}^{(1)}}(\tilde{u}_{2})r_{12}(\tilde{u}_{1}+\tilde{u}%
_{2})\stackrel{1}{K_{-}^{(1)}}(\tilde{u}_{1})r_{12}(\tilde{u}_{1}-\tilde{u}%
_{2}),\label{4b}
\end{equation}
\begin{eqnarray}
r_{12}(\tilde{u}_{2}-\tilde{u}_{1})\stackrel{1}{K_{+}^{(1)}}(\tilde{u}%
_{1})r_{12}(-\tilde{u}_{1}-\tilde{u}_{2}-2U)\stackrel{2}{K_{+}^{(1)}}(%
\tilde{u}_{2})\nonumber \\
=\stackrel{2}{K_{+}^{(1)}}(\tilde{u}_{2})r_{12}(-\tilde{u}_{1}-%
\tilde{u}_{2}-2U)\stackrel{1}{K_{+}^{(1)}}(\tilde{u}_{1})r_{12}(\tilde{u}%
_{2}-\tilde{u}_{1}).
\end{eqnarray}
For our case, the $K_{\pm }^{(1)}(u)=I$. Let us define the nested monodromy matrix 
\begin{equation}
\tilde{T}_{-}^{(1)}(\tilde{u})=T^{(1)}(\tilde{u}){T^{(1)}}^{-1}(-\tilde{u}%
)=\left( \matrix{\tilde{A}^{(1)}(\tilde{u})&\tilde{B}^{(1)}(\tilde{u})\cr
\tilde{C}^{(1)}(\tilde{u})&\tilde{D}^{(1)}(\tilde{u})\cr }\right) 
\label{4c2}
\end{equation}
which also satisfies the RE (\ref{4b}). Using the main ingredients (\ref{4a}%
)-(\ref{4c2}) describing the open BC compatible with the integrability of
the model and following all steps solving $XXZ$ open chain in \cite{SK}, one
can present the following results: 
\begin{eqnarray}
\Lambda ^{(1)}(\tilde{u},\{\tilde{u}_1\cdots \tilde{u}_M\}\{\tilde{v}_i\})\mid \Phi ^{(1)}(\tilde{u}_l,\{\tilde{v}_i\})\rangle=
\left\{\frac{2(\tilde{u}+U)}{2\tilde{u}+U}\prod ^{M}_{l=1}\frac{(\tilde{u}+\tilde{u}_l)(\tilde{u}-\tilde{u}_l-U)}
{(\tilde{u}-\tilde{u}_l)(\tilde{u}+\tilde{u}_l+U)}\right. \nonumber \\
\left.+\frac{2\tilde{u}}{2\tilde{u}+U}\prod^n_{i=1}b(\tilde{u}+\tilde{v}_i)b(\tilde{u}-\tilde{v}_i)
\prod ^{M}_{l=1}\frac{(\tilde{u}+\tilde{u}_l+2U)(\tilde{u}-\tilde{u}_l+U)}
{(\tilde{u}-\tilde{u}_l)(\tilde{u}+\tilde{u}_l+U)}\right\}\mid \Phi ^{(1)}(\tilde{u}_l,\{\tilde{v}_i\})\rangle\label{4cc}
\end{eqnarray}
provided that 
\begin{equation}
\prod_{i=1}^{n}\frac{(\tilde{u}_{j}+\tilde{v}_{i}+U)(\tilde{u}_{j}-\tilde{v}%
_{i}+U)}{(\tilde{u}_{j}+\tilde{v}_{i})(\tilde{u}_{j}-\tilde{v}_{i})}=\prod_{
\begin{array}{l}
l=1, \\ 
l\neq j
\end{array}
}^{M}\frac{(\tilde{u}_{j}+\tilde{u}_{l}+2U)(\tilde{u}_{j}-\tilde{u}_{l}+U)}{(%
\tilde{u}_{j}+\tilde{u}_{l})(\tilde{u}_{j}-\tilde{u}_{l}-U)},
\end{equation}
\[
j=1,\cdots M,
\]
which indeed ensure the cancellation of all unwanted terms in (\ref{4cc}).
Here the ''spin'' part of the multi-particle states is given by 
\begin{equation}
\mid \Phi ^{(1)}(\tilde{u}_{l},\{\tilde{v}_{i}\})\rangle=\tilde{B}^{(1)}(\tilde{u}%
_{1})\cdots \tilde{B}^{(1)}(\tilde{u}_{M})|0\rangle^{(1)},
\end{equation}
where $M$ is the number of holes with spin down, $n$ is the total number of
the holes.

Finally, if we adopt the variables $z_{\pm }(v_i)$ used in \cite{Mar}%
,i.e. 
\begin{equation}
z_-(v_i)=\frac{\cos v_i}{\sin v_i}e^{2h(v_i)},\,\,z_+(v_i)=\frac{\sin v_i}{%
\cos v_i}e^{2h(v_i)},
\end{equation}
and make a shift $\tilde{u}_j=\tilde{\lambda }_j-U/2$, the eigenvalue of the
transfer matrix (\ref{2h}) is given as

\[
\tau(u)\mid \Phi _n(v_1,\cdots,v_n)\rangle = 
\]
\begin{eqnarray}
\left\{W^+_1(u)W^-_1(u)[z_-(u)]^{2N}\prod _{i=1}^n\frac{\sin
^2u(1+z_-(v_i)/z_+(u))(1+1/z_-(v_i)z_+(u))}{\cos
^2u(1-z_-(v_i)/z_-(u))(1-1/z_-(v_i)z_-(u))}\right.  \nonumber \\
\left.+W_4^+(u)W^-_4(u)[z_+(u)]^{2N}\prod _{i=1}^n\frac{\sin
^2u(1+z_-(v_i)z_-(u))(1+z_-(u)/z_-(v_i))}{\cos
^2u(1-z_-(v_i)z_+(u))(1-z_+(u)/z_-(v_i))}\right.  \nonumber \\
\left.+W^+_2(u)W^-_2(u)\frac{2(\tilde{u}+U)}{2\tilde{u}+U}\prod _{i=1}^n%
\frac{\sin ^2u(1+z_-(v_i)/z_+(u))(1+1/z_-(v_i)z_+(u))}{\cos
^2u(1-z_-(v_i)/z_-(u))(1-1/z_-(v_i)z_-(u))}\times \right.  \nonumber \\
\left.\prod_{l=1}^M\frac{(\tilde{u}+\tilde{\lambda }_l-U/2)(\tilde{u}-\tilde{%
\lambda }_l-U/2)}{(\tilde{u}-\tilde{\lambda }_l+U/2)(\tilde{u}+\tilde{%
\lambda }_l+U/2)}\right.  \nonumber \\
\left.+W^+_2(u)W^-_2(u)\frac{2\tilde{u}}{2\tilde{u}+U}\prod _{i=1}^n\frac{%
\sin ^2u(1+z_-(v_i)z_-(u))(1+z_-(u)/z_-(v_i))}{\cos
^2u(1-z_-(v_i)z_+(u))(1-z_+(u)/z_-(v_i))}\times \right.  \nonumber \\
\left.\prod_{l=1}^M\frac{(\tilde{u}+\tilde{\lambda }_l+3U/2)(\tilde{u}-%
\tilde{\lambda }_l+3U/2)}{(\tilde{u}-\tilde{\lambda }_l+U/2)(\tilde{u}+%
\tilde{\lambda }_l+U/2)} \right\}\mid \Phi _n(v_1,\cdots,v_n)\rangle,  \label{re1}
\end{eqnarray}
provided that 
\begin{equation}
\zeta (v_i,\xi _+)\zeta (v_i,\xi _-)[z_-(v_i)]^{2N}= \prod ^M_{l=1}\frac{(%
\tilde{v}_i+\tilde{\lambda }_l-U/2)(\tilde{v}_i-\tilde{\lambda }_l-U/2)}{(%
\tilde{v}_i-\tilde{\lambda }_l+U/2)(\tilde{v}_i+\tilde{\lambda }_l+U/2)}
\label{4e}
\end{equation}
\begin{equation}
\prod_{i=1}^n\frac{(\tilde{\lambda }_j+\tilde{v}_i+U/2)(\tilde{\lambda }_j-%
\tilde{v}_i+U/2)}{(\tilde{\lambda }_j+\tilde{v}_i-U/2)(\tilde{\lambda }_j-%
\tilde{v}_i-U/2)}= \prod ^M_{
\begin{array}{l}
l=1, \\ 
l\neq j
\end{array}
}\frac{(\tilde{\lambda }_j+\tilde{\lambda }_l+U)(\tilde{\lambda }_j-\tilde{%
\lambda }_l+U)} {(\tilde{\lambda }_j+\tilde{\lambda }_l-U)(\tilde{\lambda }%
_j-\tilde{\lambda }_l-U)},  \label{4f}
\end{equation}
\[
j=1,\cdots M,\,\, i=1,\cdots ,n,
\]
where 
\begin{equation}
\zeta (u,\xi _{\pm})=\frac{e^{-h(u)}\xi _{\pm }\sin u-e^{h(u)}\cos u}{%
e^{h(u)}\xi _{\pm }\cos u-e^{-h(u)}\sin u}.
\end{equation}
If we express the variable $z_-(u_i)$ in terms of the momenta $k_i$ (hole)
by $z_-(u_i)=e^{2k_i}$, from the relation (\ref{2g}), the energy is given by 
\begin{equation}
E_n=\xi _-+\xi _+-(N/2-n)U-\sum _{i=1}^n4\cos k_i.  \label{re2}
\end{equation}
Equations (\ref{re1})-(\ref{re2}) constitute our main results of this paper.
It is found that the boundary fields are indeed nontrivial to the ground
state properties and the boundary energy of the model.

\section{Conclusion}

We have formulated the algebraic Bethe ansatz solution for the 1D Hubbard
model with open boundaries. Bethe ansatz equations, the eigenvalue of the
transfer matrix and energy spectrum have also been given. Comparing our
results with coordinate Bethe ansatz solution \cite{Coor}, the Bethe ansatz
equations (\ref{4e}) and (\ref{4f}) coincide with ones obtained in \cite
{Coor}, apart from this, we presented explicitly the eigenvalue of the
transfer matrix and conjectured the main structure of the $n$-particle
eigenvectors. The results obtained provides us with a start point to study
the thermodynamical properties of the model \cite{QISM2,FTP}. Especially,
the proposed way is available to formulate the algebraic Bethe ansatz for
other extented  Hubbard models with open BC, such as 1D Bariev open chain 
\cite{Bar,zhou1}, $U_q[Osp(2|2)]$ electronic system \cite{SL22,SLMG}, etc.
We also notice that if we add the chemical potential term $\nu \sum
^N_{j=1}\sum _s(n_{js}-1/2)$ to the Hamiltonian (\ref{2a}), the
integrability of  the model require the associated quantum $R$-matrix \cite
{GY}, which does not have crossing  unitarity, should satisfy new RE. But
the new class of the boundary $K_{\pm }$-matrices \cite{GWY} shall not
change the Bethe ansatz equations (\ref{4e}) and (\ref{4f}). Nevertheless,
if we add the Kondo impurities \cite{Kond,Kond2} $J\sum
_{ss^{^{\prime}}}a^{\dagger }_{s}{\bf \sigma }_{ss^{^{\prime}}}a_{s^{^{%
\prime}}}.{\bf S}$ to each boundaries, the model is also integrable with a
certain boundary $K_{\pm}$-matrices which lead to new Bethe ansatz
equations. We hope following this paper we shall present a class of
integrable Kondo impurities for the 1D Hubbard model in near future.\newline
\vspace{1cm}

\newpage

\begin{center}
{\bf Appendix}
\end{center}

We display the quantum ${\cal R}(u,v)$-matrix of the 1D Hubbard model below 
\cite{Wad1,Wad2} 
\[
{\small {\ \left(%
\matrix{
\rho _1&0&0&0&0&0&0&0&0&0&0&0&0&0&0&0\cr
0&-i\rho _{10}&0&0&\rho _2&0&0&0&0&0&0&0&0&0&0&0\cr 
0&0&-i\rho _{10}&0&0&0&0&0&\rho _2&0&0&0&0&0&0&0\cr
0&0&0&\rho _8&0&0&i\rho _6&0&0&-i\rho _6&0&0&\rho _3&0&0&0\cr
0&\rho _{2}&0&0&i\rho _9&0&0&0&0&0&0&0&0&0&0&0\cr 
0&0&0&0&0&-\rho _4&0&0&0&0&0&0&0&0&0&0\cr
0&0&0&i\rho  _6&0&0&-\rho _7&0&0&-\rho _5&0&0&-i\rho _6&0&0&0\cr
0&0&0&0&0&0&0&\rho _9&0&0&0&0&0&\rho _{2}&0&0\cr
0&0&\rho _{2}&0&0&0&0&0&i\rho _9&0&0&0&0&0&0&0\cr
0&0&0&-i\rho _6&0&0&-\rho _5&0&0&-\rho _7&0&0&i\rho _6&0&0&0\cr
0&0&0&0&0&0&0&0&0&0&-\rho _4&0&0&0&0&0\cr 
0&0&0&0&0&0&0&0&0&0&0&i\rho _9&0&0&\rho _2&0\cr
0&0&0&\rho _3&0&0&-i\rho _6&0&0&i\rho _6&0&0&\rho _8&0&0&0\cr
0&0&0&0&0&0&0&\rho _2&0&0&0&0&0&-i\rho _{10}&0&0\cr
0&0&0&0&0&0&0&0&0&0&0&\rho _2&0&0&-i\rho _{10}&0\cr
0&0&0&0&0&0&0&0&0&0&0&0&0&0&0&\rho _1\cr }\right)},}
\]

with the Boltzmann weights 
\begin{eqnarray}
\rho _1 & = & (\cos u~\cos v~e^{l}+\sin v~\sin u~e^{-l})\rho _2,  \nonumber
\\
\rho _4 & = & (\cos u~\cos v~e^{-l}+\sin v~\sin u~e^{l})\rho _2,  \nonumber
\\
\rho _9 & = & (\sin u~\cos v e^{-l}-\sin v~\cos u~e^{l})\rho _2,  \nonumber
\\
\rho _{10} & = & (\sin u~\cos v~e^{l}-\sin v~\cos u~e^{-l})\rho _2, 
\nonumber \\
\rho _3 & = & \frac{(\cos u~\cos v~e^{l}-\sin v~\sin u~e^{-l})}{\cos
^2u-\sin ^2v}\rho _2,  \nonumber \\
\rho _5 & = & \frac{(\cos u~\cos v~e^{-l}-\sin v~\sin u~e^{l})}{\cos
^2u-\sin ^2v}\rho _2,  \nonumber \\
\rho _6 & = & \frac{e^{-h}(\cos u~\sin u~e^{l}-\sin v~\cos v~e^{-l})}{\cos
^2u-\sin ^2v}\rho _2,  \nonumber
\end{eqnarray}
and 
\[
\rho _8=\rho _1-\rho _3;\, \rho _7= \rho _4-\rho _5,
\]
\[
l=h(u)-h(v),h=h(u)+h(v)
\]
which enjoy the following identities: 
\[
\rho _4\rho _1+\rho _9\rho _{10}=1,\, \rho _1\rho _5+\rho _3\rho _4=2,
\]
\[
\rho ^2_6=\rho _3\rho _5-1,\, \rho ^2_6=\rho _9\rho _{10}+\rho _7\rho _8.
\]

{\bf Acknowledgements} The author would like to thank M. J. Martins for
proofreading the manuscript and giving me many valuable suggestions. Many
thanks to U. Grimm, R.A. R\"omer and H.Fan for their helpful discussions in
the start of this work. This work has been supported by Fapesp (Funda\c
c\~ao de Amparo \`a Pesquisa do Estado de S\~ao Paulo).

\end{document}